\documentclass[aps,prl,twocolumn,reprint,superscriptaddress,floatfix,nofootinbib]{revtex4-2}

\usepackage{amsmath,amssymb,graphicx,bm,xcolor,dcolumn}
\usepackage{tikz}
\tikzset{
  tensor/.style        ={draw=black, line width=0.8pt},
  tensor-label/.style  ={font=\itshape\small},
  tensor-x/.style      ={tensor, fill=white},   
  tensor-m/.style      ={tensor, fill=white},   
  tensor-u/.style      ={tensor, fill=white},   
  tensor-s/.style      ={tensor, fill=white},   
  tensor-v/.style      ={tensor, fill=white},   
  tensor-r/.style      ={tensor, fill=white},   
  bond/.style          ={draw=black, line width=0.6pt},
  bond-thick/.style    ={draw=black, line width=1.2pt},
  bond-dashed/.style   ={draw=black, line width=0.6pt, dashed},
  bond-green/.style    ={draw=green!60!black, line width=0.8pt},
  bond-double/.style   ={draw=black, line width=0.5pt,
                         double distance=2.8pt},
}

\newcommand{\doublebond}[4]{%
  \pgfmathparse{abs((#1)-(#3)) < 0.01}%
  \ifnum\pgfmathresult=1
    \draw[bond] ({(#1)-1.4},{#2}) -- ({(#3)-1.4},{#4});%
    \draw[bond] ({(#1)+1.4},{#2}) -- ({(#3)+1.4},{#4});%
  \else
    \pgfmathparse{abs((#2)-(#4)) < 0.01}%
    \ifnum\pgfmathresult=1
      \draw[bond] ({#1},{(#2)-1.4}) -- ({#3},{(#4)-1.4});%
      \draw[bond] ({#1},{(#2)+1.4}) -- ({#3},{(#4)+1.4});%
    \else
      \draw[bond-double] ({#1},{#2}) -- ({#3},{#4});%
    \fi
  \fi
}

\definecolor{petroffblue}{HTML}{3F90DA}    
\definecolor{petroffgold}{HTML}{FFA90E}    
\definecolor{petroffred}{HTML}{BD1F01}     
\definecolor{petroffgrey}{HTML}{94A4A2}    
\definecolor{petroffpurple}{HTML}{832DB6}  
\definecolor{petroffbrown}{HTML}{A96B59}   
\definecolor{petroffdorange}{HTML}{E76300} 
\definecolor{petrofftan}{HTML}{B9AC70}     
\definecolor{petroffdgrey}{HTML}{717581}   
\definecolor{petroffcyan}{HTML}{92DADD}    
\definecolor{petroffsage}{HTML}{8FB496}    
\tikzset{
  tensor-c/.style = {tensor, fill=petroffblue!50},    
  tensor-t/.style = {tensor, fill=petrofftan!65},      
  tensor-a/.style = {tensor, fill=petroffgold!60},     
  tensor-p/.style = {tensor, fill=petroffpurple!40},   
  tensor-x/.style = {tensor, fill=petroffcyan!80},     
  tensor-m/.style = {tensor, fill=petroffred!35},      
  tensor-u/.style = {tensor, fill=petroffdorange!45},  
  tensor-v/.style = {tensor, fill=petroffdorange!45},  
  tensor-s/.style = {tensor, fill=petroffsage!85},     
  tensor-r/.style = {tensor, fill=petroffbrown!45},    
}
\newcolumntype{d}{D{.}{.}{-1}}
\usepackage[colorlinks=true,linkcolor=blue,citecolor=blue,urlcolor=blue]{hyperref}

\begin{document}

\title{Fast two-dimensional tensor-network contraction via subspace iteration}

\author{Yining Zhang}
\affiliation{Institute for Theoretical Physics, University of Amsterdam, Science Park 904, 1098 XH Amsterdam, The Netherlands}

\author{Philippe Corboz}
\affiliation{Institute for Theoretical Physics, University of Amsterdam, Science Park 904, 1098 XH Amsterdam, The Netherlands}

\date{\today}

\begin{abstract}
The corner transfer matrix renormalization group (CTMRG) is one of the standard
contraction methods for infinite projected entangled-pair states (iPEPS), but
its computational cost is dominated by repeated truncated singular value decompositions (SVDs). We
introduce subspace-iteration CTMRG (SI-CTMRG), a QR-based projector construction that
replaces each large-matrix SVD with an SVD of a much smaller matrix. The resulting algorithm shifts
the dominant cost from decompositions to tensor contractions, making it highly
suited to GPU acceleration and yielding speedups of up to two orders of
magnitude over standard CTMRG. We demonstrate the efficiency and accuracy of
the method for the triangular-lattice Heisenberg antiferromagnet, reaching
state-of-the-art iPEPS results on a single H100 GPU in approximately 10 hours of
computation.
\end{abstract}

\maketitle

\emph{Introduction.}---Tensor network methods have become powerful tools
for studying quantum many-body systems, providing variational
ans\"atze whose accuracy can be systematically controlled by the bond
dimension $D$. Infinite projected entangled-pair states (iPEPS)~\cite{Verstraete2004,Nishio2004,Jordan2008} provide efficient descriptions of two-dimensional quantum lattice systems and
have produced state-of-the-art results across a broad range of
strongly correlated problems, including frustrated quantum
magnets~\cite{Liao2017,Niesen2017,Jahromi2018,Hasik2021,Hasik2022,Liu2022,Jimenez2021,Lee2020,Hasik2024,Schmoll2024,Corboz2025,ZhangKitaev2025,Ma2026},
doped Hubbard and $t$-$J$ models~\cite{Corboz2014,Ponsioen2019,Ponsioen2023,ZhangtJ2026},
SU($N$) and chiral spin systems~\cite{Chen2018,Chung2019,Gauthe2020},
and fractional quantum Hall states~\cite{Weerda2024,ChenFCI2025}.

Evaluating observables on an iPEPS requires the approximate contraction of an infinite
two-dimensional tensor network, for which the corner transfer matrix
renormalization group
(CTMRG)~\cite{Nishino1996,Nishino1997,Orus2009,Corboz2011,Corboz2014,Nietner2020}
has become one of the most widely used methods, alongside algorithms such as variational uniform matrix product states (VUMPS)~\cite{ZaunerStauber2018,Fishman2018,Vanderstraeten2022} and the tensor renormalization group (TRG)~\cite{Levin2007,Xie2012} and tensor network renormalization (TNR)~\cite{Evenbly2015}. Each CTMRG iteration grows the corner
tensors by absorbing rows or columns of the network and then truncates
the resulting $\chi D^2\times\chi D^2$ enlarged corner back to
environment bond dimension $\chi$ via projectors built from
a truncated singular value decomposition (SVD), which scales as
$\mathcal{O}((\chi D^2)^3)$ with standard SVD routines and dominates the computational cost at large bond dimensions required for state-of-the-art accuracy.

Variational iPEPS optimization compounds this cost.
Gradient-based schemes~\cite{Corboz2016,Vanderstraeten2016} now
routinely rely on automatic differentiation (AD) through the CTMRG procedure~\cite{Liao2019,Hasik2021,Zhang2023,Francuz2025}, but
backpropagating through all iterations greatly increases memory use
and, together with the cubic scaling of the SVD,
restricts the bond dimensions
that can be reached in practice. Several recent works have attacked these bottlenecks:
reduced-rank and single-layer contraction schemes lower the
asymptotic scaling~\cite{Haghshenas2019,Lan2023}, split-CTMRG
variants enable simulations at larger $\chi$~\cite{Naumann2025split},
and GPU-accelerated implementations with randomized SVD (RSVD) have begun to exploit modern
hardware~\cite{Richards2025}. On the AD side, stable iterative
decompositions reduce the overall computational cost~\cite{Francuz2025}, and fixed-point differentiation removes the
need to backpropagate through every iteration~\cite{Liao2019}.
Replacing SVDs by cheaper factorizations has proved useful in
tensor-network settings, for example in QR-based MPS time evolution
relative to SVD-based implementations~\cite{Unfried2023}. More directly
for CTMRG, the QR-decomposition--based projector construction accelerates the CTMRG steps by one to two orders
of magnitude for $C_{4v}$ symmetric and $C_{3v}$ symmetric tensor networks~\cite{Zhang2025, Yang2026}.

 Building on recent developments in randomized algorithms such as RSVD~\cite{Halko2011} and the Nystr\"om method~\cite{Nystrom1930,WilliamsSeeger2001,Nakatsukasa2020}, as well as on the QR-CTMRG approach~\cite{Zhang2025}, we develop subspace-iteration CTMRG (SI-CTMRG), a CTMRG variant for generic iPEPS on square lattices. The computationally dominant SVD is replaced by QR decompositions of smaller matrices together with a small SVD, which are significantly faster and well suited to GPU acceleration. Furthermore, the new scheme naturally combines with fixed-point AD, thereby enabling highly efficient two-dimensional tensor-network calculations. On a single H100 GPU this enables fully differentiable
iPEPS optimization that reaches state-of-the-art results for the triangular Heisenberg antiferromagnet, compared to previous studies~\cite{Hasik2024,Naumann2025,NQS2026}.

\emph{iPEPS.}---An iPEPS is a variational tensor-network ansatz that
represents a two-dimensional quantum many-body state directly in the
thermodynamic limit~\cite{Verstraete2004,Nishio2004,Jordan2008}. For example, Eq.~\eqref{eq:ipeps_ansatz} shows an iPEPS ansatz built from a single rank-five, real-valued on-site
tensor $A$ repeated on a square lattice,
\begin{equation}
\resizebox{0.7\columnwidth}{!}{
\begin{tikzpicture}[x=1pt, y=-1pt]
  \tikzset{tensor-label/.style={font=\itshape\large}}
  \def\biglab{\Large}
  \def\ax{44}   
  \def\bx{26}   
  \def\by{22}   
  \def\phys{20} 
  \def\stub{16} 
  \def\Yb{100}

  \foreach \j in {0,1,2}{
    \pgfmathsetmacro\yy{\Yb-\j*\by}
    \foreach \i in {0,1}{
      \pgfmathsetmacro\xa{\i*\ax+\j*\bx}
      \pgfmathsetmacro\xb{(\i+1)*\ax+\j*\bx}
      \draw[bond] (\xa,\yy) -- (\xb,\yy);
    }
    \pgfmathsetmacro\xL{0+\j*\bx}
    \pgfmathsetmacro\xR{2*\ax+\j*\bx}
    \draw[bond-dashed] (\xL,\yy) -- (\xL-\stub,\yy);
    \draw[bond-dashed] (\xR,\yy) -- (\xR+\stub,\yy);
  }
  \foreach \i in {0,1,2}{
    \foreach \j in {0,1}{
      \pgfmathsetmacro\xa{\i*\ax+\j*\bx}     \pgfmathsetmacro\ya{\Yb-\j*\by}
      \pgfmathsetmacro\xb{\i*\ax+(\j+1)*\bx} \pgfmathsetmacro\yb{\Yb-(\j+1)*\by}
      \draw[bond] (\xa,\ya) -- (\xb,\yb);
    }
    \pgfmathsetmacro\xt{\i*\ax+2*\bx}  \pgfmathsetmacro\yt{\Yb-2*\by}
    \pgfmathsetmacro\xs{\xt+0.7071*\stub} \pgfmathsetmacro\ys{\yt-0.6*\stub}
    \draw[bond-dashed] (\xt,\yt) -- (\xs,\ys);
    \pgfmathsetmacro\xbm{\i*\ax}       \pgfmathsetmacro\ybm{\Yb}
    \pgfmathsetmacro\xbs{\xbm-0.7071*\stub} \pgfmathsetmacro\ybs{\ybm+0.6*\stub}
    \draw[bond-dashed] (\xbm,\ybm) -- (\xbs,\ybs);
  }
  \foreach \i in {0,1,2}{
    \foreach \j in {0,1,2}{
      \pgfmathsetmacro\xx{\i*\ax+\j*\bx}
      \pgfmathsetmacro\yy{\Yb-\j*\by}
      \draw[bond] (\xx,\yy) -- (\xx,\yy+\phys);
    }
  }
  \foreach \i in {0,1,2}{
    \foreach \j in {0,1,2}{
      \pgfmathsetmacro\xx{\i*\ax+\j*\bx}
      \pgfmathsetmacro\yy{\Yb-\j*\by}
      \filldraw[tensor-a] (\xx,\yy) circle (7pt);
    }
  }
  \node[font=\biglab] at (-58,\Yb-1*\by) {$|\Psi\rangle \;\approx$};
  \pgfmathsetmacro\xc{1*\ax+1*\bx} \pgfmathsetmacro\yc{\Yb-1*\by}
  \node[tensor-label] at (\xc-5,\yc-14) {$A$};
  \pgfmathsetmacro\xd{1.5*\ax+2*\bx} \pgfmathsetmacro\yd{\Yb-2*\by}
  \node[tensor-label] at (\xd,\yd-9) {$D$};
\end{tikzpicture}}
\label{eq:ipeps_ansatz}
\end{equation}
where each $A$ carries one physical index of dimension $d$, spanning the
local Hilbert space of a lattice site, and four virtual indices of bond
dimension $D$ connecting to the neighboring tensors, such that $A$ has in total
$dD^{4}$ parameters for a dense tensor without exploiting symmetries. Contracting each tensor $A$ with its conjugate
$A^{\dagger}$ along the physical index yields the double-leg tensor $a$,
\begin{equation}
\resizebox{0.55\columnwidth}{!}{
\begin{tikzpicture}[x=1pt, y=-1pt]
  \tikzset{tensor-label/.style={font=\itshape\large}}
  \def\biglab{\Large}
  \def\cx{0}
  \def\Atop{32}      
  \def\Abot{62}      
  \def\mid{47}       
  \def\rr{7pt}
  \draw[bond] (\cx,\Atop+7) -- (\cx,\Abot-7);
  \draw[bond] (\cx+7,\Atop) -- (\cx+34,\Atop);   
  \draw[bond] (\cx+7,\Abot) -- (\cx+34,\Abot);
  \draw[bond] (\cx-7,\Atop) -- (\cx-34,\Atop);   
  \draw[bond] (\cx-7,\Abot) -- (\cx-34,\Abot);
  \draw[bond] (\cx+4.3,\Atop-5.5) -- (\cx+18.6,\Atop-23.4);  
  \draw[bond] (\cx+4.3,\Abot-5.5) -- (\cx+18.6,\Abot-23.4);
  \draw[bond] (\cx-4.3,\Atop+5.5) -- (\cx-18.6,\Atop+23.4);  
  \draw[bond] (\cx-4.3,\Abot+5.5) -- (\cx-18.6,\Abot+23.4);
  \filldraw[tensor-a] (\cx,\Atop) circle (\rr);
  \filldraw[tensor-a] (\cx,\Abot) circle (\rr);
  \node[tensor-label] at (\cx-14,\Atop-15) {$A$};
  \node[tensor-label] at (\cx-25,\Abot+13) {$A^{\dagger}$};
  \node[font=\biglab] at (74,\mid) {$=$};
  \def\dx{128}
  \filldraw[tensor-a] (\dx,\mid) circle (\rr);
  \node[tensor-label] at (\dx+14,\mid-11) {$a$};
  \doublebond{\dx+7}{\mid}{\dx+27}{\mid}
  \doublebond{\dx-7}{\mid}{\dx-27}{\mid}
  \doublebond{\dx}{\mid-7}{\dx}{\mid-27}
  \doublebond{\dx}{\mid+7}{\dx}{\mid+27}
\end{tikzpicture}}
\label{eq:double_layer}
\end{equation}
and the resulting square-lattice network of
$a$ tensors represents the norm $\langle\Psi|\Psi\rangle$. Expectation
values of local observables are obtained by contracting the
same network with the corresponding operators through physical indices.

\begin{figure}[t]
\centering
\resizebox{0.95\columnwidth}{!}{\input{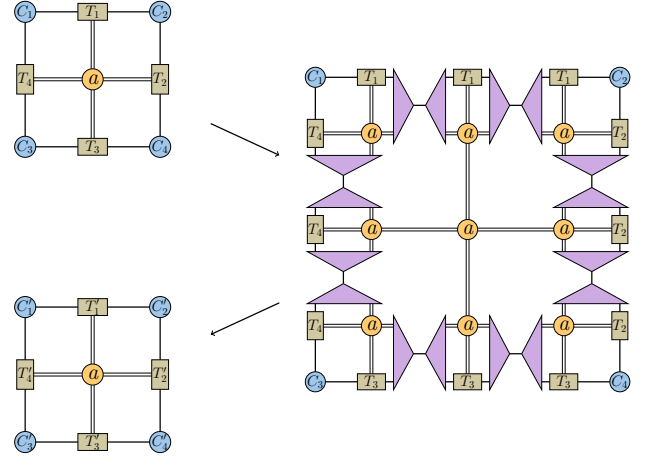}}
\caption{\label{fig:ctmrg-step}
One full CTMRG step. The single-site
environment $(C_{1,\ldots,4},T_{1,\ldots,4})$ (top
left) grows in all directions (right). Each
corner becomes an enlarged corner, and each edge is enlarged by
inserting the bulk tensor $a$. Projectors are constructed to truncate
the $\chi D^{2}$ bonds back to $\chi$. Contracting the enlarged corners
and edges yields the updated environment $(C_{1,\ldots,4}',T_{1,\ldots,4}')$ (bottom left).}
\end{figure}

\emph{Standard CTMRG.}---For an infinite two-dimensional tensor network,
CTMRG provides a systematic approximation to the contraction required
for computing physical observables. Starting from an initial
environment described by boundary tensors $C$ and $T$, the method grows
the environment in all directions by absorbing rows and columns of the
network. Each growth step produces enlarged boundary tensors, for
which the environment bond dimension $\chi$ increases by a factor of
$D^2$. A renormalization step then truncates these enlarged tensors
back to bond dimension $\chi$, as illustrated in
Fig.~\ref{fig:ctmrg-step}. Each CTMRG step therefore updates the environment to represent a larger system. As the environment converges, the tensors $C$ and $T$ represent, respectively, the corners and half-rows (or half-columns) of the system in the thermodynamic limit. The accuracy of this approximation is controlled by the bond dimension $\chi$.

The dominant computational cost lies in constructing the projectors used for renormalization. For each direction, a pair of projectors (a ``right'' projector $P_r$ renormalizing the large $\chi D^2$ open legs on the right and a ``left'' projector $P_l$ acting on the left $\chi D^2$ open legs) is constructed to truncate the enlarged environment from $\chi D^2$ back to $\chi$. The standard ``half-system'' approach contracts the four enlarged corners $\tilde C_1, \tilde C_2, \tilde C_3, \tilde C_4$ along their $\chi D^2$ bonds into the effective environment $M \in\mathbb{R}^{N\times N}$ of the virtual bonds to be truncated, with $N=\chi D^2$,
\begin{equation}
\resizebox{0.85\columnwidth}{!}{
%
%
%
%
\begin{tikzpicture}[x=1pt, y=-1pt]
  \tikzset{
    tensor-label/.append style = {font=\itshape\Large},
  }
  \begin{scope}[shift={(0, 0)}]
    \filldraw[tensor-a] (128, 112) circle (10pt);
    \node[tensor-label] at (128, 112) {a};
    \filldraw[tensor-t] (56, 98)   rectangle (72, 126);
    \node[tensor-label] at (64, 112)  {$T_4$};
    \filldraw[tensor-t] (114, 40)  rectangle (142, 56);
    \node[tensor-label] at (128, 48)  {$T_1$};
    \filldraw[tensor-c] (64, 48)   circle (10pt);
    \node[tensor-label] at (64, 48)   {$C_1$};
    \filldraw[tensor-t] (56, 178)  rectangle (72, 206);
    \node[tensor-label] at (64, 192)  {$T_4$};
    \filldraw[tensor-a] (128, 192) circle (10pt);
    \node[tensor-label] at (128, 192) {a};
    \filldraw[tensor-a] (192, 112) circle (10pt);
    \node[tensor-label] at (192, 112) {a};
    \filldraw[tensor-a] (192, 192) circle (10pt);
    \node[tensor-label] at (192, 192) {a};
    \filldraw[tensor-c] (64, 256)  circle (10pt);
    \node[tensor-label] at (64, 256)  {$C_3$};
    \filldraw[tensor-t] (114, 248) rectangle (142, 264);
    \node[tensor-label] at (128, 256) {$T_3$};
    \filldraw[tensor-t] (178, 248) rectangle (206, 264);
    \node[tensor-label] at (192, 256) {$T_3$};
    \filldraw[tensor-t] (248, 178) rectangle (264, 206);
    \node[tensor-label] at (256, 192) {$T_2$};
    \filldraw[tensor-t] (248, 98)  rectangle (264, 126);
    \node[tensor-label] at (256, 112) {$T_2$};
    \filldraw[tensor-t] (178, 40)  rectangle (206, 56);
    \node[tensor-label] at (192, 48)  {$T_1$};
    \filldraw[tensor-c] (256, 48)  circle (10pt);
    \node[tensor-label] at (256, 48)  {$C_2$};
    \filldraw[tensor-c] (256, 256) circle (10pt);
    \node[tensor-label] at (256, 256) {$C_4$};
    \draw[bond-thick]  (64, 58)   -- (64, 98);
    \draw[bond-thick]  (74, 48)   -- (114, 48);
    \doublebond{128}{56}{128}{102}
    \doublebond{72}{112}{118}{112}
    \doublebond{138}{192}{182}{192}
    \doublebond{138}{112}{152}{112}
    \doublebond{182}{112}{168}{112}
    \doublebond{72}{192}{118}{192}
    \doublebond{192}{122}{192}{182}
    \draw[bond-thick]  (64, 206)  -- (64, 246);
    \draw[bond-thick]  (74, 256)  -- (114, 256);
    \draw[bond-thick]  (142, 256) -- (178, 256);
    \doublebond{128}{202}{128}{248}
    \doublebond{192}{202}{192}{248}
    \doublebond{202}{192}{248}{192}
    \doublebond{202}{112}{248}{112}
    \draw[bond-thick]  (256, 58)  -- (256, 98);
    \doublebond{192}{56}{192}{102}
    \draw[bond-thick]  (142, 48)  -- (154, 48);
    \draw[bond-thick]  (178, 48)  -- (166, 48);
    \draw[bond-thick]  (206, 48)  -- (246, 48);
    \draw[bond-thick]  (256, 126) -- (256, 178);
    \draw[bond-thick]  (206, 256) -- (246, 256);
    \draw[bond-thick]  (256, 206) -- (256, 246);
    \draw[bond-thick]        (64, 126)  -- (64, 178);
    \doublebond{128}{122}{128}{182}
  \end{scope}

  \node[font=\Huge] at (305, 152) {$=$};

  \begin{scope}[shift={(113.8, -166.3)}, scale=2.2]
    \filldraw[tensor-c, rounded corners=11.75pt] (107.3, 98.0) rectangle (134.8, 127.0);
    \node[tensor-label] at (121.05, 112.5) {$\tilde C_1$};
    \filldraw[tensor-c, rounded corners=11.75pt] (171.0, 98.0) rectangle (198.5, 127.0);
    \node[tensor-label] at (184.75, 112.5) {$\tilde C_2$};
    \filldraw[tensor-c, rounded corners=11.75pt] (107.3, 162.0) rectangle (134.8, 191.0);
    \node[tensor-label] at (121.05, 176.5) {$\tilde C_4$};
    \filldraw[tensor-c, rounded corners=11.75pt] (171.0, 162.0) rectangle (198.5, 191.0);
    \node[tensor-label] at (184.75, 176.5) {$\tilde C_3$};
    \draw[bond-thick] (134.8, 105.0) -- (147.0, 105.0);
    \draw[bond-thick] (171.0, 105.0) -- (159.0, 105.0);
    \doublebond{134.8}{120.0}{147.0}{120.0}
    \doublebond{171.0}{120.0}{159.0}{120.0}
    \doublebond{134.8}{169.0}{171.0}{169.0}
    \draw[bond-thick] (134.8, 184.0) -- (171.0, 184.0);
    \doublebond{177.25}{162.0}{177.25}{127.0}
    \draw[bond-thick] (192.25, 162.0) -- (192.25, 127.0);
    \doublebond{128.55}{162.0}{128.55}{127.0}
    \draw[bond-thick] (113.55, 162.0) -- (113.55, 127.0);
  \end{scope}
\end{tikzpicture}}.
\label{eq:rho_M_diag}
\end{equation}
An SVD is then performed and truncated to bond dimension $\chi$, giving $M\approx U\,S\,V^{\dagger}$. The projectors are then constructed as~\cite{Corboz2014,Wang2011,Huang2012}:
\begin{equation}
\resizebox{0.8\columnwidth}{!}{
%
\begin{tikzpicture}[x=1pt, y=-1pt]
  \tikzset{
    tensor-label/.append style = {font=\itshape\Large},
  }
  \filldraw[tensor-c, rounded corners=9.75pt] (12.5, 38.25) rectangle (51.5, 89.75);
  \node[tensor-label] at (32, 64) {$\tilde C_2$};

  \filldraw[tensor-c, rounded corners=9.75pt] (76.5, 38.25) rectangle (115.5, 89.75);
  \node[tensor-label] at (96, 64) {$\tilde C_3$};

  \filldraw[tensor-v] (158, 64) -- (130, 32) -- (130, 96) -- cycle;
  \node[tensor-label] at (144, 64) {$V$};

  \filldraw[tensor-s] (185, 64) circle (8pt);
  \node[tensor-label, anchor=north] at (185, 76) {$s^{-1/2}$};

  \draw[bond] (12.5, 48) -- (0, 48);
  \doublebond{12.5}{80}{0}{80}

  \draw[bond] (51.5, 48) -- (76.5, 48);
  \doublebond{51.5}{80}{76.5}{80}

  \draw[bond] (115.5, 48) -- (130, 48);
  \doublebond{115.5}{80}{130}{80}

  \draw[bond] (158, 64) -- (177, 64);
  \draw[bond] (193, 64) -- (214, 64);

  \node[tensor-label] at (240, 64) {$=$};
  \filldraw[tensor-p] (300, 64) -- (272, 32) -- (272, 96) -- cycle;
  \node[tensor-label] at (286, 64) {$P_l$};
  \draw[bond] (272, 48) -- (260, 48);
  \doublebond{272}{80}{260}{80}
  \draw[bond] (300, 64) -- (312, 64);

  \begin{scope}[xshift=40pt]
  \filldraw[tensor-s] (87, 176) circle (8pt);
  \node[tensor-label, anchor=north] at (87, 188) {$s^{-1/2}$};

  \filldraw[tensor-u] (114, 176) -- (142, 144) -- (142, 208) -- cycle;
  \node[tensor-label] at (128, 176) {$U^{\dagger}$};

  \filldraw[tensor-c, rounded corners=9.75pt] (156.5, 150.25) rectangle (195.5, 201.75);
  \node[tensor-label] at (176, 176) {$\tilde C_4$};

  \filldraw[tensor-c, rounded corners=9.75pt] (220.5, 150.25) rectangle (259.5, 201.75);
  \node[tensor-label] at (240, 176) {$\tilde C_1$};

  \draw[bond] (95, 176) -- (114, 176);
  \draw[bond] (142, 160) -- (156.5, 160);
  \doublebond{142}{192}{156.5}{192}
  \draw[bond] (195.5, 160) -- (220.5, 160);
  \doublebond{195.5}{192}{220.5}{192}
  \draw[bond] (259.5, 160) -- (272, 160);
  \doublebond{259.5}{192}{272}{192}
  \draw[bond] (58, 176) -- (79, 176);

  \filldraw[tensor-p] (-28, 176) -- (0, 144) -- (0, 208) -- cycle;
  \node[tensor-label] at (-14, 176) {$P_r$};
  \draw[bond] (-28, 176) -- (-40, 176);
  \draw[bond] (0, 160) -- (12, 160);
  \doublebond{0}{192}{12}{192}
  \node[tensor-label] at (32, 176) {$=$};
  \end{scope}
\end{tikzpicture}}
\label{eq:svd_projectors_diag}
\end{equation}
The computational cost of this standard SVD is $\mathcal{O}(\chi^3 D^6)$, which is the computational bottleneck of the standard CTMRG approach.

\emph{QR-based projector.}---
For the $C_{4v}$-symmetric case, the QR-CTMRG algorithm~\cite{Zhang2025} proposes a contraction scheme combining QR decomposition and CTMRG. With a QR-based projector, QR-CTMRG can be viewed as fusing the QR algorithm for eigendecomposition~\cite{GolubVanLoan2013} into the CTMRG steps: in the $C_{4v}$-symmetric case, the repeated QR iterations drive the corner tensor to its eigendecomposition. Since the QR algorithm can be interpreted as an implicit power method, the $C_{4v}$-symmetric case demonstrates the effectiveness of incorporating power-method updates for decompositions directly into the CTMRG steps. The eigenvectors obtained with the QR-based method eventually converge to the same fixed point as those obtained from the standard exact eigendecomposition at each CTMRG step. Based on this observation, we propose to use the QR-based method to construct projectors for CTMRG steps in generic iPEPS settings. Recent developments in randomized algorithms, such as RSVD and the Nystr\"om method~\cite{Halko2011,Nystrom1930,WilliamsSeeger2001,Nakatsukasa2020}, have shown the effectiveness of randomized algorithms for low-rank approximation of large matrices. The accuracy of these methods is mainly controlled by the oversampling parameter $p$ and the number of subspace iterations. The QR-based method builds on these ideas and incorporates the subspace iterations and oversampling parameter $p$ into the CTMRG steps. The QR-based method is implemented as follows:

1. At each CTMRG step, one subspace-iteration update refines the isometries (rangefinders) $X_{n}$, $Y_{n}$ toward the dominant $(\chi+p)$-dimensional subspaces of $M$ (the effective environment of the virtual bonds, as shown in Eq.~\eqref{eq:rho_M_diag}), where the oversampling parameter $p$ denotes a small number of additional environment states. The isometries are initialized with random Gaussian matrices, and at step $n$ the update is

\begin{equation}
\resizebox{0.69\columnwidth}{!}{
%
%
\begin{tikzpicture}[x=1pt, y=-1pt]
  \tikzset{tensor-label/.append style = {font=\itshape\large}}

  \filldraw[tensor-x] (32, 32) -- (4, 0) -- (4, 64) -- cycle;
  \node[tensor-label] at (16, 32) {$X_{n}$};
  \draw[bond] (4, 16) -- (-12, 16);              
  \doublebond{4}{48}{-12}{48}                    
  \draw[bond] (32, 32) -- (44, 32);              
  \filldraw[tensor-r] (53, 32) circle (9);
  \node[tensor-label] at (53, 32) {$R_{x}$};
  \draw[bond] (62, 32) -- (74, 32);              
  \node[tensor-label, font=\Large] at (93, 32) {$=$};
  \filldraw[tensor-m, rounded corners=9.75pt] (122, 6.25) rectangle (161, 57.75);
  \node[tensor-label] at (141.5, 32) {$M^{T}$};
  \draw[bond] (122, 16) -- (112, 16);             
  \doublebond{122}{48}{112}{48}                   
  \filldraw[tensor-m, rounded corners=9.75pt] (173, 6.25) rectangle (212, 57.75);
  \node[tensor-label] at (192.5, 32) {$M$};
  \draw[bond] (161, 16) -- (173, 16);            
  \doublebond{161}{48}{173}{48}                  
  \filldraw[tensor-x] (252, 32) -- (224, 0) -- (224, 64) -- cycle;
  \node[tensor-label] at (236, 32) {$X_{\scriptscriptstyle n{-}1}$};
  \draw[bond] (212, 16) -- (224, 16);            
  \doublebond{212}{48}{224}{48}                  
  \draw[bond] (252, 32) -- (264, 32);            

  \filldraw[tensor-r] (17, 126) circle (9);
  \node[tensor-label] at (17, 126) {$L_{y}$};
  \draw[bond] (8, 126) -- (-12, 126);            
  \draw[bond] (26, 126) -- (38, 126);            
  \filldraw[tensor-x] (38, 126) -- (66, 94) -- (66, 158) -- cycle;
  \node[tensor-label] at (52, 126) {$Y_{n}$};
  \draw[bond] (66, 110) -- (76, 110);            
  \doublebond{66}{142}{76}{142}                  
  \node[tensor-label, font=\Large] at (98, 126) {$=$};
  \filldraw[tensor-x] (130, 126) -- (158, 94) -- (158, 158) -- cycle;
  \node[tensor-label] at (144, 126) {$Y_{\scriptscriptstyle n{-}1}$};
  \draw[bond] (130, 126) -- (120, 126);           
  \draw[bond] (158, 110) -- (170, 110);          
  \doublebond{158}{142}{170}{142}                
  \filldraw[tensor-m, rounded corners=9.75pt] (170, 100.25) rectangle (209, 151.75);
  \node[tensor-label] at (189.5, 126) {$M$};
  \draw[bond] (209, 110) -- (221, 110);          
  \doublebond{209}{142}{221}{142}                
  \filldraw[tensor-m, rounded corners=9.75pt] (221, 100.25) rectangle (260, 151.75);
  \node[tensor-label] at (240.5, 126) {$M^{T}$};
  \draw[bond] (260, 110) -- (272, 110);          
  \doublebond{260}{142}{272}{142}                
\end{tikzpicture}}
\label{eq:xy_iter}
\end{equation}
such that after the power update, a QR decomposition (or LQ decomposition) is performed separately, obtaining $X_{n}$ and $Y_{n}$ for orthonormalization. Toward the end of the CTMRG iterations, $X_{n}$ and $Y_{n}$ converge to the leading $(\chi+p)$-dimensional subspaces corresponding to $U$ and $V$ in the SVD of $M$.

2. The SVD used to compute the projectors is therefore performed only on the small $(\chi+p)\times(\chi+p)$ matrix $\rho = Y_n\,M\,X_n = u\,s\,v^{\dagger}$, and is truncated to $\chi$ singular values, i.e., $\rho\approx u_\chi\,s_\chi\,v^{\dagger}_\chi$,
\begin{equation}
\resizebox{0.9\columnwidth}{!}{
%
%
%
\begin{tikzpicture}[x=1pt, y=-1pt]
  \tikzset{
    tensor-label/.append style = {font=\itshape\Large},
    tensor-labelsvd/.append style = {font=\itshape\Large},
  }

  \begin{scope}[shift={(-96, -144)}]
    \filldraw[tensor-x] (114, 176) -- (142, 144) -- (142, 208) -- cycle;
    \node[tensor-label] at (128, 176) {$Y_n$};
    \filldraw[tensor-m, rounded corners=13.5pt]
             (153.6, 149) rectangle (230.4, 203);
    \node[tensor-label] at (192, 176) {$M$};
    \filldraw[tensor-x] (270, 176) -- (242, 144) -- (242, 208) -- cycle;
    \node[tensor-label] at (256, 176) {$X_n$};
    \draw[bond] (114, 176) -- (96, 176);               
    \draw[bond] (142, 162.5) -- (153.6, 162.5);        
    \doublebond{142}{189.5}{153.6}{189.5}              
    \draw[bond] (230.4, 162.5) -- (242, 162.5);        
    \doublebond{230.4}{189.5}{242}{189.5}              
    \draw[bond] (270, 176) -- (288, 176);              
  \end{scope}

  \node[font=\Large] at (216, 32) {$=$};

  \begin{scope}[shift={(112, -240)}]
    \filldraw[tensor-u] (151, 263) rectangle (169, 281);
    \node[tensor-labelsvd] at (160, 272) {$u$};
    \filldraw[tensor-s] (192, 272) circle (8pt);
    \node[tensor-labelsvd] at (192, 272) {$s$};
    \filldraw[tensor-v] (215, 263) rectangle (233, 281);
    \node[tensor-labelsvd] at (224, 272) {$v^{\dagger}$};
    \draw[bond] (169, 272) -- (184, 272);
    \draw[bond] (200, 272) -- (215, 272);
    \draw[bond] (151, 272) -- (128, 272);
    \draw[bond] (233, 272) -- (256, 272);
  \end{scope}

\end{tikzpicture}}
\label{eq:rho_hat_diag}
\end{equation}

3. The projectors for the CTMRG step are reassembled with:

\begin{equation}
\resizebox{0.82\columnwidth}{!}{
%
%
\begin{tikzpicture}[x=1pt, y=-1pt]
  \tikzset{
    tensor-label/.append style = {font=\itshape\Large},
  }
  \filldraw[tensor-c, rounded corners=9.75pt] (12.5, 38.25) rectangle (51.5, 89.75);
  \node[tensor-label] at (32, 64) {$\tilde C_2$};

  \filldraw[tensor-c, rounded corners=9.75pt] (76.5, 38.25) rectangle (115.5, 89.75);
  \node[tensor-label] at (96, 64) {$\tilde C_3$};

  \filldraw[tensor-x] (158, 64) -- (130, 32) -- (130, 96) -- cycle;
  \node[tensor-label] at (144, 64) {$X_n$};

  \filldraw[tensor-v] (167, 55) rectangle (185, 73);
  \node[tensor-label] at (176, 64) {$v_{\chi}$};

  \filldraw[tensor-s] (208, 64) circle (8pt);
  \node[tensor-label, anchor=north] at (208, 76) {$s_{\chi}^{-1/2}$};

  \draw[bond] (12.5, 48) -- (0, 48);
  \doublebond{12.5}{80}{0}{80}

  \draw[bond] (51.5, 48) -- (76.5, 48);
  \doublebond{51.5}{80}{76.5}{80}

  \draw[bond] (115.5, 48) -- (130, 48);
  \doublebond{115.5}{80}{130}{80}

  \draw[bond] (158, 64) -- (167, 64);
  \draw[bond] (185, 64) -- (200, 64);
  \draw[bond] (216, 64) -- (224, 64);

  \node[tensor-label] at (240, 64) {$=$};
  \filldraw[tensor-p] (300, 64) -- (272, 32) -- (272, 96) -- cycle;
  \node[tensor-label] at (286, 64) {$P_l$};
  \draw[bond] (272, 48) -- (260, 48);        
  \doublebond{272}{80}{260}{80}              
  \draw[bond] (300, 64) -- (312, 64);        

  \begin{scope}[xshift=40pt]
  \filldraw[tensor-s] (64.4, 176) circle (8pt);
  \node[tensor-label, anchor=north] at (64.4, 188) {$s_{\chi}^{-1/2}$};

  \filldraw[tensor-u] (87, 167) rectangle (105, 185);
  \node[tensor-label] at (96, 176) {$u_{\chi}^{\dagger}$};

  \filldraw[tensor-x] (114, 176) -- (142, 144) -- (142, 208) -- cycle;
  \node[tensor-label] at (128, 176) {$Y_n$};

  \filldraw[tensor-c, rounded corners=9.75pt] (156.5, 150.25) rectangle (195.5, 201.75);
  \node[tensor-label] at (176, 176) {$\tilde C_4$};

  \filldraw[tensor-c, rounded corners=9.75pt] (220.5, 150.25) rectangle (259.5, 201.75);
  \node[tensor-label] at (240, 176) {$\tilde C_1$};

  \draw[bond] (72.4, 176) -- (87, 176);
  \draw[bond] (105, 176) -- (114, 176);
  \draw[bond] (142, 160) -- (156.5, 160);
  \doublebond{142}{192}{156.5}{192}
  \draw[bond] (195.5, 160) -- (220.5, 160);
  \doublebond{195.5}{192}{220.5}{192}
  \draw[bond] (259.5, 160) -- (272, 160);
  \doublebond{259.5}{192}{272}{192}
  \draw[bond] (48, 176) -- (56.4, 176);

  \filldraw[tensor-p] (-28, 176) -- (0, 144) -- (0, 208) -- cycle;
  \node[tensor-label] at (-14, 176) {$P_r$};
  \draw[bond] (-28, 176) -- (-40, 176);      
  \draw[bond] (0, 160) -- (12, 160);         
  \doublebond{0}{192}{12}{192}               
  \node[tensor-label] at (32, 176) {$=$};
  \end{scope}
\end{tikzpicture}}
\label{eq:projectors_diag}
\end{equation}
where the updated isometries are applied. The $(\chi+p)\times(\chi+p)$ unitary matrices $u,v$ from the small SVD of $\rho$ fix the basis within the $(\chi+p)$-dimensional subspaces represented by $X_n$ and $Y_n$. They are absorbed back into the updated isometries to be used in the next CTMRG step,
\begin{equation}
\resizebox{0.95\columnwidth}{!}{
%
%
%
\begin{tikzpicture}[x=1pt, y=-1pt]
  \tikzset{
    tensor-label/.append style = {font=\itshape\Large},
  }

  \filldraw[tensor-x] (4, 32) -- (32, 0) -- (32, 64) -- cycle;
  \node[tensor-label] at (19, 32) {$Y_{n}$};
  \draw[bond] (4, 32)   -- (-20, 32);            
  \draw[bond] (32, 16)  -- (52, 16);             
  \doublebond{32}{48}{52}{48}                    
  \node[tensor-label] at (64, 32) {$\leftarrow$};
  \filldraw[tensor-u] (84, 23) rectangle (102, 41);
  \node[tensor-label] at (93, 32) {$u^{\dagger}$};
  \draw[bond] (84, 32)  -- (76, 32);             
  \filldraw[tensor-x] (111, 32) -- (139, 0) -- (139, 64) -- cycle;
  \node[tensor-label] at (126, 32) {$Y_{n}$};
  \draw[bond] (102, 32) -- (111, 32);            
  \draw[bond] (139, 16) -- (156, 16);            
  \doublebond{139}{48}{156}{48}                  

  \node[tensor-label] at (170, 32) {$,$};

  \filldraw[tensor-x] (212, 32) -- (184, 0) -- (184, 64) -- cycle;
  \node[tensor-label] at (197, 32) {$X_{n}$};
  \draw[bond] (184, 16) -- (172, 16);            
  \doublebond{184}{48}{172}{48}                  
  \draw[bond] (212, 32) -- (220, 32);            
  \node[tensor-label] at (232, 32) {$\leftarrow$};
  \filldraw[tensor-x] (288, 32) -- (260, 0) -- (260, 64) -- cycle;
  \node[tensor-label] at (273, 32) {$X_{n}$};
  \draw[bond] (260, 16) -- (244, 16);            
  \doublebond{260}{48}{244}{48}                  
  \filldraw[tensor-v] (297, 23) rectangle (315, 41);
  \node[tensor-label] at (306, 32) {$v$};
  \draw[bond] (288, 32) -- (297, 32);            
  \draw[bond] (315, 32) -- (335, 32);            
\end{tikzpicture}}
\label{eq:xy_recycle}
\end{equation}

4. With the gauge-fixing step in Eq.~\eqref{eq:xy_recycle}
and sign-fixing of the SVD~\cite{BroAcarKolda2008}, CTMRG approaches a fixed point at which $C$
and $T$ converge element-wise. Consequently, the dominant subspaces also
stabilize, which allows us to recycle $X_n$ and $Y_n$ across CTMRG
iterations. In practice, after the first $\sim\!5$ warm-up steps following step 1, the
recycled bases rarely need updating for the accuracy concerned, so projector construction is
dominated by forming the  products $\tilde C_2\,\tilde C_3\,X_n$ and
$Y_n\,\tilde C_4\,\tilde C_1$.
The dominant cost of projector construction is then the tensor contraction, which scales as $\mathcal{O}(\chi^3 D^4)$, while the QR decomposition and the small SVD contribute only $\mathcal{O}(\chi^3 D^2)$ and $\mathcal{O}(\chi^3)$, respectively. By contrast, in the standard approach the full SVD scales as $\mathcal{O}(\chi^3 D^6)$. Although iterative solvers can reduce this scaling to $\mathcal{O}(\chi^3 D^4)$, the large prefactor still makes the decomposition step expensive, particularly on GPUs. Consequently, the QR-based approach is substantially faster, especially on GPUs.

\emph{Variational optimization.}---The variational parameters in iPEPS can be optimized either by imaginary-time evolution~\cite{Jiang2008,Phien2015,Czarnik2019} or by minimizing the variational energy~\cite{Corboz2016,Vanderstraeten2016}. Combining energy minimization with automatic differentiation to compute the energy gradients is currently the most accurate approach~\cite{Liao2019,Xie2020,Hasik2021,Zhang2023}. In automatic differentiation frameworks, the forward computation is recorded or traced, allowing gradients with respect to the input parameters to be computed by applying the chain rule in reverse mode. Because all of the linear-algebra operations are differentiable, SI-CTMRG can naturally be combined with the AD framework. We use the limited-memory Broyden--Fletcher--Goldfarb--Shanno (L-BFGS) algorithm for optimization. The CTMRG environment, together with the $X$ and $Y$ isometries, is recycled between optimization iterations, except during line-search steps. As the optimization converges, the recycled $X$ and $Y$ also converge faster in their relevant subspaces during the CTMRG steps, further accelerating the optimization.
Typically, at least three CTMRG steps with subspace iteration are performed, regardless of the subspace convergence of $X$ and $Y$.

\emph{Differentiable fixed point.}---CTMRG defines a map
$(C,T)\mapsto\mathcal{F}(C,T,A)$. At the fixed point, the CTM environment $x\equiv(C^\star,T^\star)$ satisfies $x=\mathcal{F}(x(A),A)$, and observables take the form $\mathcal{O}(A)=g(A,C^\star,T^\star)$. Rather
than unrolling 10--30 CTMRG steps through AD, the fixed-point equation is differentiated via the implicit function theorem,
\begin{equation}
\frac{\partial\mathcal{O}}{\partial A}
= \frac{\partial g}{\partial A}
+ \frac{\partial g}{\partial(C,T)}\!
\Bigl[I-\tfrac{\partial\mathcal{F}}{\partial(C,T)}\Bigr]^{\!-1}\!
\frac{\partial\mathcal{F}}{\partial A},
\label{eq:implicit}
\end{equation}
which is solved by backpropagating through a fixed-point iteration, and the basic operation involves just the vector-Jacobian products with the single-step iteration function~\cite{Liao2019}.
The peak memory is that of a single CTMRG step and is therefore
independent of the number of forward iterations.

\emph{Triangular Heisenberg benchmark.}---To test the proposed method, we consider the challenging spin-$1/2$ nearest-neighbor Heisenberg antiferromagnet on the triangular lattice ~\cite{Capriotti1999,WhiteChernyshev2007}, with Hamiltonian
\begin{equation}
   H=J\sum_{\langle i,j\rangle}\bm{S}_i\!\cdot\!\bm{S}_j.
\end{equation}
The ground state is known to exhibit $120^\circ$ long-range magnetic order, in which the spins lie in a plane with an angle of $120^\circ$ between the spin directions on the three sublattices. In order to apply the CTMRG described above, the triangular lattice is treated as a square lattice with an additional diagonal interaction~\cite{Bauer2012,Niesen2018}.
Using a spiral iPEPS ansatz~\cite{Hasik2024}, the $120^\circ$ ordered ground state with  wave vector $k = (2\pi/3, 2\pi/3)$ can be encoded using a single-site square-lattice iPEPS ansatz.

\begin{figure}[t]
\includegraphics[width=\columnwidth]{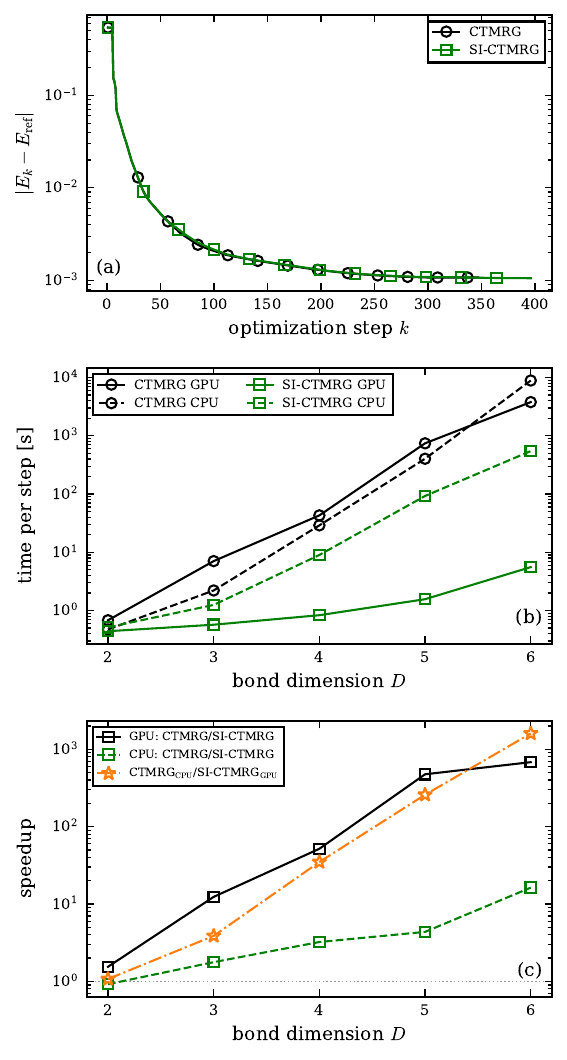}
\caption{\label{fig:timing}
CTMRG vs.\ SI-CTMRG for the triangular Heisenberg model. (a)~Convergence of
the energy, $|E_k{-}E_{\mathrm{ref}}|$ per L-BFGS iteration at $D{=}5$,
$\chi{=}100$, where $E_{\mathrm{ref}}=-0.55160$ is the thermodynamic-limit reference energy of Ref.~\cite{Hasik2024}. 
(b)~Time per optimization step
(forward$+$backward) vs.\ $D$ at $\chi{=}20,45,80,100,144$ for
$D{=}2,\ldots,6$ on a single H100 GPU (solid lines) and an 8-core
CPU (dashed lines).
(c)~Speedup ratios: algorithmic
($\text{CTMRG}/\text{SI-CTMRG}$ at fixed hardware) on GPU and on CPU, and
the combined ratio $\text{CTMRG}_{\mathrm{CPU}}/\text{SI-CTMRG}_{\mathrm{GPU}}$.}
\end{figure}

We first benchmark the optimization of a $D=5$ spiral iPEPS.
Figure~\ref{fig:timing}(a) compares energy convergence vs.\ optimization steps at $\chi=100$ using the standard CTMRG approach and
SI-CTMRG\@. SI-CTMRG runs use an
oversampling parameter $p=5$. In the subspace-iteration update, we start recycling $X_n$ and $Y_n$ once the corresponding dominant $\chi$-dimensional subspaces of successive steps are sufficiently aligned, i.e. $\overline{\sin^2\theta}\equiv\frac{1}{\chi}\sum_{i=1}^{\chi}\sin^2\theta_i<10^{-3}$, where $\theta_i$ are the principal angles between the current and previous subspaces. In the standard approach, gradients are obtained by
unrolling the full computational graph in the backward pass, whereas in
SI-CTMRG they are computed using fixed-point differentiation. The two
methods converge in a very similar manner, demonstrating the accuracy
of the new approach.

Figures~\ref{fig:timing}(b) and~\ref{fig:timing}(c) show the time per optimization step and
the corresponding speedups as functions of the bond dimension $D$. For fair comparison of the timings, instead of using fixed-point differentiation, for both standard CTMRG and SI-CTMRG, we unroll the full computational graph in the backward pass. The benchmarks were performed on an 8-core CPU and an H100 GPU\@. At $D=6$,
SI-CTMRG is about 16 times faster on the CPU and up to 680 times
faster on the GPU\@. Comparing CPU timings for the standard scheme with
GPU timings for the new scheme shows an overall speedup of up to
1600. The main source of this improvement is that, in the standard
scheme, about $99\%$ of the computational time is spent on the SVD
decomposition. In the new scheme, the SVD is performed only on a small
$\chi\times\chi$ matrix, with cost $\mathcal{O}(\chi^3)$, and becomes
negligible at large bond dimension. Among the decompositions, the QR decomposition holds the dominant complexity, which scales as $\mathcal{O}(\chi^3 D^2)$, though the computational costs of the QR and the SVD are comparable in the actual implementation. At $D=8,\chi=224$, together the
decompositions account for less than $2\%$ of the total run time on the GPU.
As a result, the overall
computational bottleneck shifts from the decomposition to the
tensor-network contraction, which scales as $\mathcal{O}(\chi^3 D^4)$
and is well suited to GPU acceleration.

We next push the SI-CTMRG computations to large bond dimensions up to $D=8$, and extrapolate the data for the ground state energy and order parameter to the exact infinite bond dimension $D\!\to\!\infty$ limit. We perform the latter based on finite correlation length scaling (FCLS)~\cite{RamsCorboz2018,Vanhecke2019,Hasik2024}, which is similar to conventional finite-size scaling, but where the system size is replaced by the iPEPS correlation length $\xi^*(D)$ induced by the finite bond dimension. To improve the FCLS data, we implement the diagonal mirror symmetry of the $120^\circ$ ordered state in the local $A$ tensor.

\begin{table}[!tb]
\caption{\label{tab:energies}
Ground-state data per bond dimension, with SI-CTMRG and
diagonal-reflection-symmetric iPEPS. $\chi_{\mathrm{opt}}$ is the
optimization environment bond dimension; $\xi(\chi_{\mathrm{opt}})$ is the
correlation length at $\chi_{\mathrm{opt}}$ from the
transfer-matrix spectrum; $\xi^{*}\equiv\lim_{\chi\to\infty}\xi(D,\chi)$
is obtained using the method of
Ref.~\cite{RamsCorboz2018}. $e$ and $m$ are the energy per site
and local magnetic moment, measured at the largest $\chi$ of
a post-optimization $\chi$-ramp ($\chi_{\mathrm{max}}=800$ for
$D=4,\ldots,7$ and $\chi_{\mathrm{max}}=640$ for $D=8$);
 $(\xi^{*}, e, m)$ are the input to the
FCLS extrapolation.}
\begin{ruledtabular}
\begin{tabular}{cccddd}
$D$ & $\chi_{\mathrm{opt}}$ &
\multicolumn{1}{c}{$\xi(\chi_{\mathrm{opt}})$} &
\multicolumn{1}{c}{$\xi^{*}$} &
\multicolumn{1}{c}{$e$ (per site)} &
\multicolumn{1}{c}{$m$}\\
\hline
4 &  80 & 1.3245 & 1.4619 & -0.548182 & 0.27727 \\
5 & 100 & 1.7282 & 2.0036 & -0.550314 & 0.24893 \\
6 & 144 & 2.0743 & 2.4576 & -0.551063 & 0.23198 \\
7 & 196 & 2.5340 & 3.0047 & -0.551443 & 0.21992 \\
8 & 224 & 2.9748 & 3.6920 & -0.551640 & 0.21108 \\
\end{tabular}
\end{ruledtabular}
\end{table}

\begin{figure}[t]
\centering
\includegraphics[width=\columnwidth]{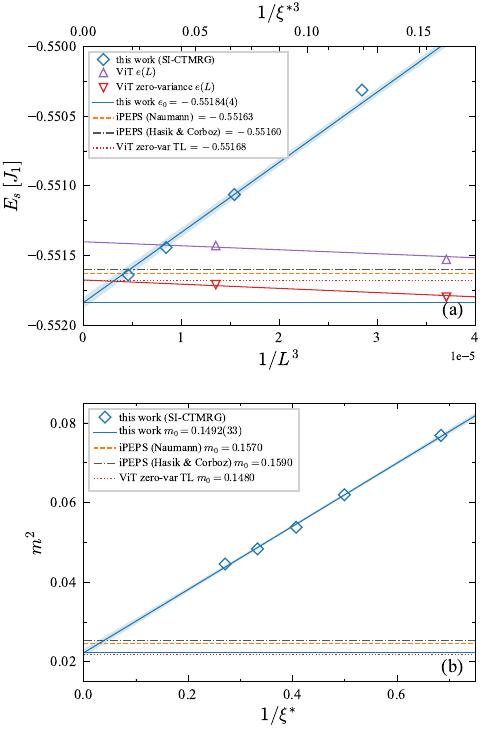}
\caption{\label{fig:fcls}
Triangular-lattice Heisenberg benchmark.
(a) Energy per site of our $D=4,\dots,8$ optimized states plotted
against $1/\xi^{*3}$ (top axis) together with the ViT wave function energies and its corresponding
zero-variance extrapolated points of Ref.~\cite{NQS2026} against $1/L^{3}$
(bottom axis); the solid blue line is our linear FCLS extrapolation
giving $e_0=-0.55184(4)$. Horizontal lines show other thermodynamic-limit (TL)
reference values: iPEPS results by Naumann \emph{et al.}~\cite{Naumann2025}
($-0.55163$), Hasik \& Corboz~\cite{Hasik2024} ($-0.55160$), and
the ViT zero-variance result ~\cite{NQS2026} ($-0.55168$).
(b) Squared sublattice magnetization $m^{2}$ vs.\ $1/\xi^{*}$
for the same states; the solid blue line shows the linear extrapolation
$m^{2}(\xi^{*})=m_0^{2}+c/\xi^{*}$ giving $m_0=0.1492(33)$. Horizontal
lines indicate the corresponding squared reference magnetizations, with
$m_0=0.157$ from Naumann \emph{et al.}, $m_0=0.159$ from
Hasik \& Corboz, and $m_0=0.148$ from the ViT zero-variance
extrapolation.}
\end{figure}

The ground state is optimized using $D$-ramping, where we start from $D=4$ and initialize each larger-$D$ optimization from the converged result at the previous, smaller $D$. For each optimized state we run a post-optimization $\chi$-ramp, evaluating the forward CTMRG at a sequence of environment dimensions up to $\chi_{\mathrm{max}}=800$ ($\chi_{\mathrm{max}}=640$ for $D=8$), with the oversampling parameter $p$ chosen to be $15\%$ of the environment dimension $\chi$. We extract $\xi(D,\chi)$ from the eigenvalues of the transfer matrix and extrapolate it to $\chi\!\to\!\infty$ following Ref.~\cite{RamsCorboz2018}; we denote this extrapolated correlation length by $\xi^{*}(D)\equiv\lim_{\chi\to\infty}\xi(D,\chi)$. The energy $e$ and local magnetic moment $m=\sqrt{\langle S_x\rangle^2+\langle S_y\rangle^2+\langle S_z\rangle^2}$ are taken at $\chi_{\mathrm{max}}$, where they are sufficiently converged.

The results  $(\xi^{*},e,m)$ for $D=4,\dots,8$ are listed in Table~\ref{tab:energies}. As shown in Fig.~\ref{fig:fcls}, a linear energy extrapolation $e(\xi^{*})=e_0+a/\xi^{*3}$ over $D=4$--$8$ yields
\begin{equation}
e_0=-0.55184(4),\qquad m_0 = 0.1492(33),
\end{equation}
where the magnetization is obtained from a linear fit $m^2(\xi^{*})=m_0^2+c/\xi^{*}$. The quoted uncertainty is the standard error of the extrapolated value. Our extrapolated energy lies ${\sim}1.6\times 10^{-4}$ below the vision-transformer (ViT) neural-network quantum-state (NQS) result
$e_0^{\mathrm{NQS}}=-0.55168(2)$ of Ref.~\cite{NQS2026}. Our magnetization $m_0=0.1492(33)$ agrees, within one standard
deviation, with the NQS value $m_0^{\mathrm{NQS}}=0.148(1)$~\cite{NQS2026} and lies
roughly $6\%$ below the iPEPS value $m_0^{\mathrm{HC}}=0.159(2)$ of Ref.~\cite{Hasik2024}. The total optimization time for obtaining the iPEPS tensor was around 10 hours on a single H100 GPU.




\emph{Discussion.}---
In this work we introduced SI-CTMRG, a QR-based CTMRG scheme with subspace
iteration and demonstrated its efficiency and its potential for
obtaining state-of-the-art iPEPS results. In this scheme,
the dominant computational cost shifts
from decompositions to contractions, thereby making the method
particularly well suited to GPU acceleration. Furthermore, the method naturally combines with fixed-point
differentiation, reducing memory usage and improving computational
efficiency. We have shown that, even for the triangular
Heisenberg model, state-of-the-art results can be obtained within a
few hours on a single H100 GPU, highlighting the potential of this
approach for other challenging problems. Although the main text focuses
on a single-site iPEPS ansatz, the present scheme generalizes naturally
to multi-site $L_x\times L_y$ unit cells, as discussed in the
End Matter.  Likewise, while the present calculations use dense tensors without exploiting global symmetries such as
$U(1)$~\cite{Singh2011,Bauer2011}, the extension of the present contraction scheme to symmetric
tensor networks is a natural direction for future work. Thus, our work establishes SI-CTMRG as a powerful method for efficient two-dimensional tensor-network calculations for increasingly complex models.

\begin{acknowledgments}
\emph{Acknowledgments}
We thank Qi Yang and Juraj Hasik for discussions.
This project has received funding from the European Research Council
(ERC) under the European Union's Horizon 2020 research and innovation
programme (grant agreement No.~101001604).
This work was carried out on the Dutch national e-infrastructure
with the support of SURF Cooperative.
An implementation of the algorithm is available online~\cite{Repo}.
\end{acknowledgments}

\bibliographystyle{apsrev4-2}
\bibliography{references}

\onecolumngrid
\vspace{1.5\baselineskip}
\begin{center}
\textbf{End Matter}
\end{center}
\vspace{0.5\baselineskip}
\twocolumngrid

\emph{Larger unit-cell CTMRG.}---While in the main text we focused on the single-site iPEPS ansatz, SI-CTMRG can be generalized to arbitrary unit cells of size $L_x\times L_y$ that are tiled periodically over the infinite two-dimensional lattice.

The main difference from the single-site case is that the projector construction and renormalization are now performed column by column across the unit cell ~\cite{Corboz2011}. For example, the left-move update of the corner tensors $C^{[x,y]}_{1}$ is done by contracting $C^{[x-1,y]}_{1}$, $T^{[x,y]}_{1}$ and the projector $\tilde P^{[x-1,y]}$, constructed in a similar way as in the single-site case; the analogous contractions for $T^{\prime\,[x,y]}_4$ and $C^{\prime\,[x,y]}_4$ involve the corresponding $P^{[x-1,y-1]}$ and $\tilde P^{[x-1,y]}$ projectors. The QR-based method for projector construction is therefore applied naturally to the larger unit-cell case, with separate $X^{[x,y]}_n$ and $Y^{[x,y]}_n$ isometries for each site in the unit cell. Unlike in the single-site case, initial warm-up procedures are required to initialize the environment tensors before applying power iterations for updating the initial random Gaussian matrices $X^{[x,y]}_0$ and $Y^{[x,y]}_0$, to ensure convergence to the correct leading subspaces. The warm-up step only needs to be performed once at the very beginning of the optimization, and instead of the full SVD, RSVD can be applied. Therefore, the warm-up stage only takes a very small fraction of the total optimization time.
\begin{figure}[!ht]
\centering
\includegraphics[width=\columnwidth]{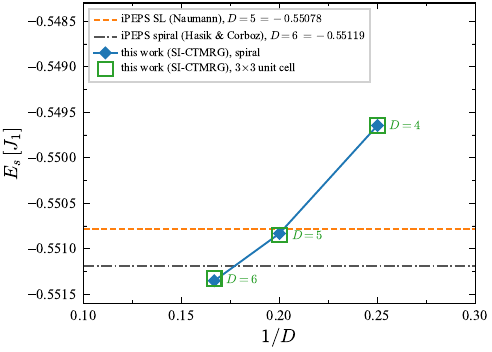}
\caption{\label{fig:triangular_heisenberg_unit_cell}
Ground-state energy per site $E_s$ of the triangular Heisenberg
antiferromagnet versus $1/D$, obtained with SI-CTMRG. Blue diamonds show
the single-site spiral ansatz and green open squares the three-sublattice
$3\times 3$ unit-cell ansatz ($D=4,5,6$); the two ans\"atze yield matching
energies at each $D$. Horizontal lines mark the $D=5$ square-lattice iPEPS result
of Naumann \emph{et al.}~\cite{Naumann2025} ($-0.55078$) and the $D=6$ spiral
iPEPS result of Hasik \& Corboz~\cite{Hasik2024} ($-0.55119$). The $3\times 3$ unit-cell energies are optimized at $\chi_{\mathrm{opt}}=80,100,144$ and
evaluated up to $\chi_{\mathrm{max}}=800,800,656$ for $D=4,5,6$,
respectively.}
\end{figure}

To test the effectiveness of SI-CTMRG for larger unit cells, we also performed calculations of the $120^\circ$ ordered triangular Heisenberg antiferromagnet using a three-sublattice ansatz with three distinct on-site tensors realizing the $120^\circ$ order in a $3\times 3$ unit cell. As in the single-site case, we use L-BFGS and fixed-point AD for the optimization. The resulting ground-state energies for $D=4,5,6$ are evaluated with sufficiently large $\chi$ to ensure convergence. Figure~\ref{fig:triangular_heisenberg_unit_cell} compares the large unit-cell ansatz to the single-site spiral ansatz and shows that the two ansätze yield matching optimized ground-state energies at each $D$.

\end{document}